\documentclass[pre,aps,11pt]{revtex4-1}

\usepackage[utf8]{inputenc} 
\usepackage{latexsym,amsmath,verbatim}
\usepackage{mathrsfs}
\usepackage{mathtools}
\DeclarePairedDelimiter\ceil{\lceil}{\rceil}
\DeclarePairedDelimiter\floor{\lfloor}{\rfloor}
\usepackage{hyperref}
\usepackage{color,soul}
\usepackage{graphicx}

\definecolor{orange}{rgb}{1,0.5,0}

\newcommand{\bistate}[2]{\text{#1}\lhd\text{#2}}

\newcommand{\csentence}[1]{\textbf{#1}}

\renewcommand{\[}{\left[}
\renewcommand{\]}{\right]}

\begin{document}

\title{Assessing reliable human mobility patterns from higher-order memory in mobile communications}

\author{Manlio De Domenico}
\author{Joan T. Matamalas}
\author{Alex Arenas}
\affiliation{Departament d'Enginyeria Inform\`{a}tica i Matem\`{a}tiques, Universitat Rovira i Virgili, 43007 Tarragona, Spain}

\begin{abstract} 
Understanding how people move within a geographic area, e.g. a city, a country or the whole world, is fundamental in several applications, from predicting the spatio-temporal evolution of an epidemics to inferring migration patterns. Mobile phone records provide an excellent proxy of human mobility, showing that movements exhibit a high level of memory.
However, the precise role of memory in widely adopted proxies of mobility, as mobile phone records, is unknown. Here we use 560 millions of call detail records from Senegal to show that standard Markovian approaches, including higher-order ones, fail in capturing real mobility patterns and introduce spurious movements never observed in reality. We introduce an adaptive memory-driven approach to overcome such issues. At variance with Markovian models, it is able to realistically model conditional waiting times, i.e. the probability to stay in a specific area depending on individual's historical movements.
Our results demonstrate that in standard mobility models the individuals tend to diffuse faster than what observed in reality, whereas the predictions of the adaptive memory approach significantly agree with observations. We show that, as a consequence, the incidence and the geographic spread of a disease could be inadequately estimated when standard approaches are used, with crucial implications on resources deployment and policy making during an epidemic outbreak.

\end{abstract}

\maketitle

\section{Introduction}
People move following complex dynamical patterns at different geographical scales, e.g. among areas of the same city, among cities and regions of the same country or among different countries. Such patterns have been recently revealed by using human mobility proxies~\cite{gonzalez2008understanding,balcan2009multiscale,song2010limits,simini2012universal,lima2016routing} and, intriguingly, some specific patterns tend to repeat more than others, with evidences \cite{balcan2011phase,rosvall2014memory} of memory of meaningful locations playing a fundamental role in our understanding of human mobility. In fact, human dynamics might significantly affect how epidemics spread \cite{ferguson2005strategies,colizza2007reaction,balcan2009multiscale,balcan2011phase,wesolowski2012quantifying} or how people migrate from one country to another~\cite{simini2012universal}.

The collaboration between researchers and mobile operators recently opened new promising directions to gather information about human movements, country demographics and health, faster and cheaper than before \cite{eagle2006reality,gonzalez2008understanding,eagle2009inferring,wesolowski2012quantifying,deville2014dynamic,amini2014impact,tizzoni2014mobility,wesolowski2014commentary,calabrese2014urban,blondel2015survey}. In fact, mobile phones heterogeneously penetrated both rural and urban communities, regardless of richness, age or gender, providing evidences that mobile technologies can be used to
obtain real-time information about individual's location and social activity, in order to build realistic demographics and socio-economics maps of a whole country \cite{wesolowski2012heterogeneous}. Mobile data have been successfully used in a wide variety of applications, e.g., to estimate population densities and their evolution at national scales \cite{deville2014dynamic}, to confirm social theories of behavioral adaptation \cite{eagle2009community} and to capture anomalous behavioral patterns associated to religious, catastrophic or massive social events \cite{dobra2014spatiotemporal}. Even more recently, the public availability of mobile phone data sets further revolutionized the field, e.g., by allowing ubiquitous sensing to map poverty, to monitor social segregation and to optimize information campaigns to reduce epidemics spreading \cite{amini2014impact,lima2015exploiting}, to cite just some of them \cite{blondel2015survey}.

Although some limitations, mobile phone data still provide one the most powerful tools for sensing complex social systems and represent a valuable proxy for studies where human mobility plays a crucial role \cite{ferguson2005strategies,colizza2007reaction,colizza2008epidemic,gonzalez2008understanding,balcan2009multiscale,song2010limits,balcan2011phase,simini2012universal,tatem2012mapping,wesolowski2012quantifying,lima2015exploiting,tizzoni2014mobility}. Milestone works in this direction have shown that human trajectories exhibit more temporal and spatial regularity than previously thought. Individuals tend to return to a few highly frequented locations and to follow simple reproducible patterns \cite{gonzalez2008understanding,lima2016routing}, allowing a higher accuracy in predicting their movements \cite{song2010limits} and significantly affecting the spreading of transmittable diseases \cite{balcan2011phase}. However, the increasing interest for using mobile phone data in applications should be accompanied by a wise usage of the information they carry on. In fact, an inadequate model accompanied by incomplete data and scarce knowledge of other fundamental factors influencing the model itself, might lead, for instance, to a wrong estimation of the incidence of an epidemics and its evolution \cite{pandey2014strategies}.

Here, we used high-quality mobile phone data, consisting of more than 560 millions of call detail records, to show that standard approaches might significantly overestimate mobility transitions between distinct geographical areas, making difficult to build a realistic model of human mobility. To overcome this issue, we developed an adaptive memory-driven model based on empirical observations that better captures existing correlations in human dynamics, showing that it is more suitable than classical memoryless or higher-order models to understand how individuals move and, for instance, might spread a disease.

\section{Materials and Methods}
\subsection{Markovian model of human mobility}

Let us consider a physical mobility network composed by nodes, representing geographic areas, connected by weighted edges, representing the fraction of individual movements among them. Usually, the weights are inferred from geolocated activities of individuals, e.g. the consecutive airports where a plane departs and lands or, as in this work, the cell towers where a person makes consecutive calls.

A standard approach to deal with mobility models of dynamics \cite{song2010limits,simini2012universal,balcan2011phase,wesolowski2012quantifying,wesolowski2013use,tizzoni2014mobility,wesolowski2014commentary} is to consider each node as a state of a Markov process, obtaining the flux between any pair of nodes from consecutive calls, and to build a mobility matrix $F_{ij}$ encoding the probability that an individual in node $i$ will move to node $j$ ($i,j=1,2,...,n$). Here, we use a similar approach to build the mobility matrix for each individual $\ell=1, 2, ..., \mathcal{L}$ separately and we then average over the whole set of mobility matrices, to obtain the transition probability of an individual, on average:
\begin{eqnarray}
F_{ij}=\frac{\sum\limits_{\ell=1}^{\mathcal{L}} f^{(\ell)}_{ij}}{\sum\limits_{\ell=1}^{\mathcal{L}} \sum\limits_{k=1}^{n} f^{(\ell)}_{ik}},
\end{eqnarray}
where $f^{(\ell)}_{ij}$ is the number of times the individual $\ell$ makes at least one call in node $j$ after making at least one call in node $i$.

We did not impose a specific time window to calculate transitions, to avoid introducing biases and undesired effects due to the choice of the temporal range and it is worth remarking that other normalizations can be considered depending on data and metadata availability \cite{tizzoni2014mobility}. Where not otherwise specified, we considered the mobility matrix obtained from the whole period of observation. This model is known as ``first-order'' (or 1-memory) because the present state is the only information required to choose the next state. Although very useful, this has the fundamental disadvantage that it does not account for mobility memory. In fact, it is very likely that an individual moves to a neighboring area (by means of a car or public transportation) to work and after a few hours he or she will go back to the original position. This effect has been shown to be relevant, for instance, at country level, where individuals fly from one city to another and often go back to their origin instead of moving towards a different city \cite{rosvall2014memory}. This memory is an intrinsic property of human mobility and must be taken into account for a realistic modeling of people movements between different geographic areas. When memory is taken into account, each physical node (e.g., $i\in \mathcal{A}$) is replaced by the corresponding state-nodes (e.g., $\bistate{i}{j}\in \tilde{\mathcal{A}}$ if memory is of order 2) encoding the information that an individual is in node $i$ when he or she comes from $j$. While $\mathbf{F}$ encodes information about the network of $n$ physical nodes, we need to introduce a new matrix $\mathbf{H}$ to encode information about the network of $n^{2}$ state-nodes, accounting for the allowed binary combinations (e.g. $\bistate{k}{j}$, $j,k=1,2,...,n$) between physical nodes. Similarly, higher-order memory can be taken into account by building appropriate matrices.

We use different mobility matrices to build different mobility models. Let $N_{i}(t)$ indicate the population of the physical node $i\in \mathcal{A}$ at time $t$, then the $n$ mobility equations describing how the flux of people diffuses through the network are given by
\begin{eqnarray}
N_{i}(t+1) = \sum_{j=1}^{n} F_{ji} N_{j}(t).
\end{eqnarray}
In the case of $\tau-$memory, we indicate by $\tilde{N}_{\alpha}(t)$ the population of the state-node $\alpha \in \tilde{A}$ at time $t$ and the $n^{\tau}$ mobility equations required to describe the same process are given by
\begin{eqnarray}
\tilde{N}_{\alpha}(t+1) = \sum_{\rho=1}^{n^{\tau}} H_{\rho\alpha} \tilde{N}_{\rho}(t).
\end{eqnarray}
The population in each physical node at time $t$ is given by the sum of the population in the corresponding state-nodes. It is worth remarking that, in general, the matrix $\mathbf{H}$ can be a function of time as well and the equations would keep their structural form.

\subsection{Adaptive memory model of human mobility}

However, spatial human mobility is quite complex and (higher-order) Markovian dynamics might no be suitable to model peculiar patterns such as returning visits and conditional waiting times, i.e. the probability to stay in a location depending on the origin of the travel.

We will discuss better this point in the following. Let us consider, for instance, the call sequence \textit{BBBBCCCSSS} made by an individual traveling between three American cities: Chicago, Boston and San Antonio. The main drawbacks of Markovian models  -- of order lower than three -- become evident in a scenario like this one, because the number of consecutive calls in the same city exceeds the memory of the model and the spatial information about previously visited locations is lost. Clearly, in presence of more complicated patterns, increasing the order of the model will not solve the issue and some information will be inevitably lost. Alternatively, we could aggregate consecutive calls in the same place to a single identifier, e.g. the previous sequence would be reduced to \textit{BCS}. In this case, a Markovian model would preserve the spatial information and correctly identify the transitions between the three cities, at the price of losing information about how many calls have been made in each place.

\begin{figure}[ht!]
        \includegraphics[width=0.6\textwidth]{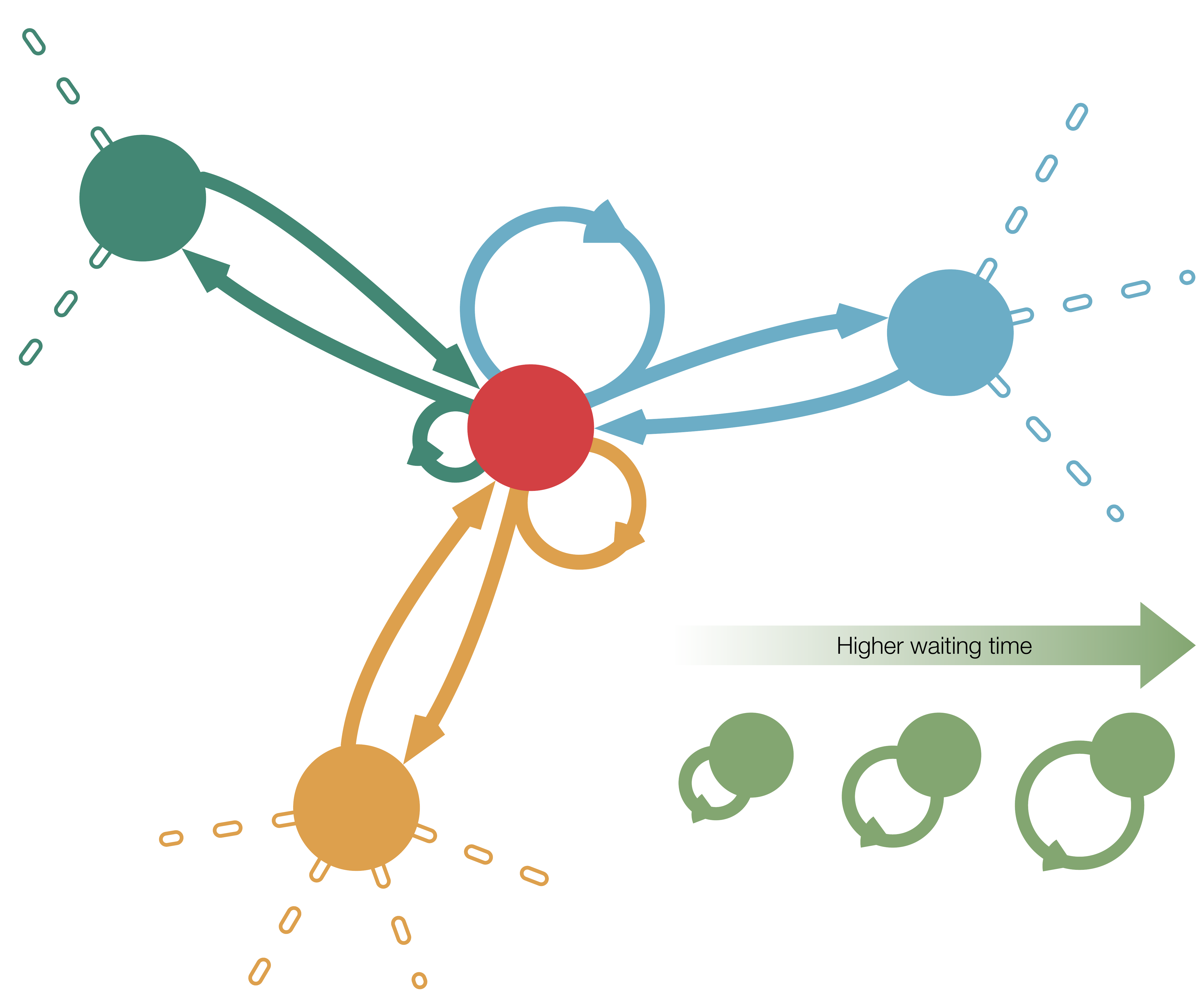}
        \caption{\label{fig:introProblem}\csentence{Conditional waiting times.} An example of human mobility between four different places. Individuals from green, blue and orange nodes move to the red central node and, after some time, go back to their previous location. The amount of time spent in the red node by individuals coming from the other nodes depends on their previous location, and it is represented by self-loops of different size.}
\end{figure}

In absence of detailed temporal information about calling activity, the number of consecutive calls in a specific location can be used as a proxy: higher the number of calls larger the waiting time. The temporal information about the amount of time spent in each location is critical for many dynamical processes like spreading or congestion. We assert that this time, like the next visited location, is conditioned by previous movements of the individuals. To illustrate this, we use the example shown in Fig.~\ref{fig:introProblem}, where people from three different places (nodes blue, green and orange) go to the same destination (node red), stay some time in there, and come back to the origin of their trip. The self-loops in the central (red) node represent the time spent there, the color encoding individuals coming from different origins and the size encoding the amount of time spent. For instance, individuals coming from the blue node wait more than individuals coming from the green node. This type of dependence is what we call conditional waiting time.

To better appreciate this fact, let us consider holiday trips. Individuals making expensive intercontinental trips tend to spend more time visiting the destination than individuals making cheaper trips, achieving a good trade-off between the travel cost and the time spent. Another emblematic case is urban mobility. For instance, the red node might be an expensive commercial area, the green node a wealthy neighborhood and the blue node a less wealthy area. In this scenario, individuals coming from the less wealthy area are more likely to be qualified workers in the commercial one, with long visits that in turn are not frequent. Conversely, individuals coming from the wealthy neighborhood are more likely to make frequent, although shorter, visits for shopping, for instance.

The importance of accounting for conditional waiting times will be evident later, when we will consider the spreading of an epidemics in a country.

\begin{figure*}[ht!]
        \includegraphics[width=\textwidth]{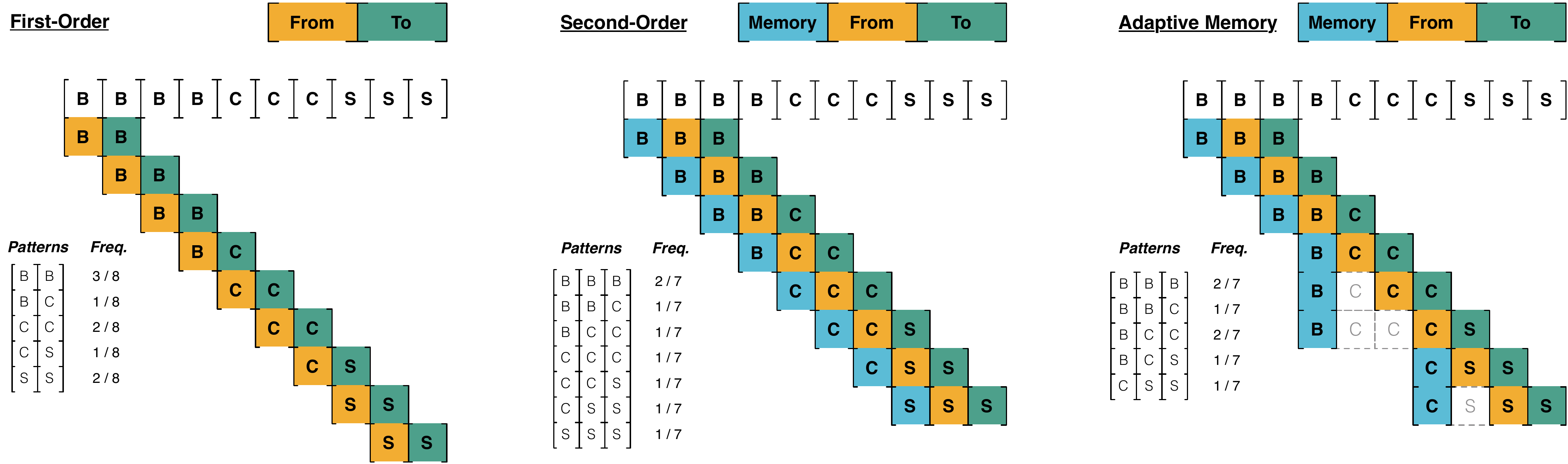}
        \caption{\label{fig:toy-model}\csentence{Comparing different mobility models}. Mobility models built from a representative sequence of mobile phone calls (BBBBCCCSSS) made, for instance, by an individual during travels between three American cities, namely Chicago (C), San Antonio (S) and Boston (B). Let us focus on the pattern ${S}\leftarrow{C}\leftarrow{B}$, that is the real sequence of movements in the geographical space. The first-order model predicts a probability of $\frac{1}{64}$, the second-order model a probability of $\frac{1}{49}$, whereas the adaptive 2-memory estimates a probability of $\frac{1}{7}$, closer to observation.}
\end{figure*}

Here, we propose a mobility model, that we name \emph{adaptive memory}, able to account for conditional waiting times. At first order, the method is equivalent to a classical first-order Markovian model, whereas significant differences emerge for increasing memory with respect to standard approaches. For instance, at second order, the 2-memory mobility matrix is built between all possible pairs of nodes (2-states), as in a standard second-order Markovian model. However, instead of considering transitions between areas in the sequence of calls, as a second-order Markovian model does, transitions in the sequence of distinct geographical areas are considered. This point is crucial, and we better clarify it with the example shown in Fig.\,\ref{fig:toy-model}, where the differences between adaptive memory and Markovian models, in terms of probability assigned to different mobility patterns, are reported.

\begin{figure*}[ht!]
  \centering
        \includegraphics[width=1\textwidth]{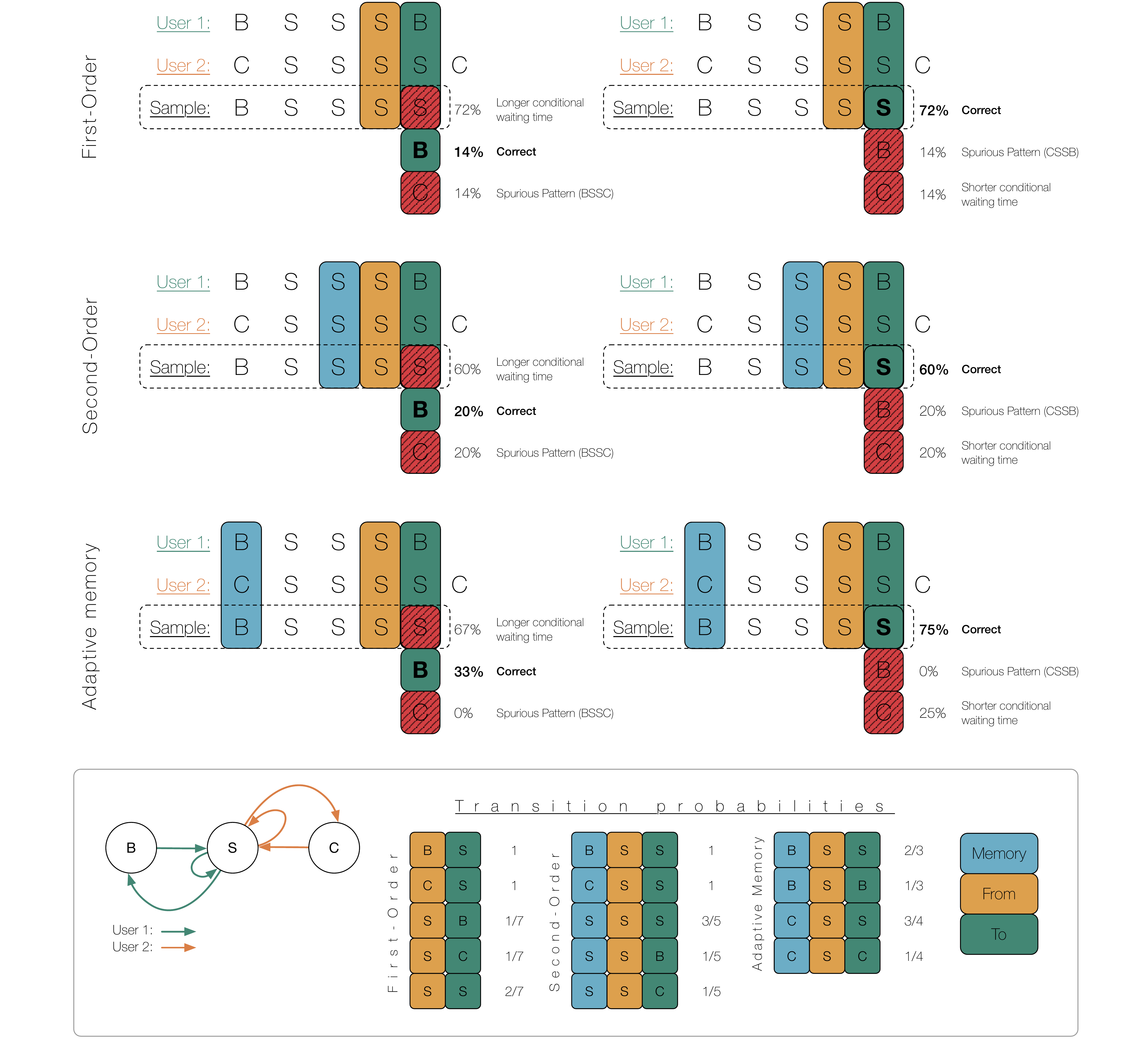}
        \caption{\label{fig:TwoSequence}\csentence{Predicting individual mobility.} Using the sequence of calls made by two different users (1 and 2) -- starting from two different locations (B and C) and visiting a new location S -- we build first-order and second-order Markov models, as well as the adaptive memory one. We use each mobility model to generate the possible mobility sequences. Given that there are two empirical starting points, we originated the sampled sequences in B and C, respectively. In the figure, for each sample, we report the fraction of times it is reproducing observation (``Correct''), it is a non-observed mobility pattern (``Spurious Pattern'') and it is underestimating or overestimating waiting times (``Longer/Shorter Conditional Waiting Time'').}
\end{figure*}

The importance of such differences is reflected in the ability of each model to predict successive individual movements. In fact, the presence of spurious or under-represented patterns might significantly affect the results, as shown in Fig.~\ref{fig:TwoSequence}. In this example, two sequences of phone calls generated by two different users moving between three cities -- B, C and S -- are considered. Markovian models generate spurious patterns that are never observed in the data, issue not affecting the adaptive memory model by construction. Morover, our approach predicts the next movement with more accuracy than Markovian ones, because it correctly takes into account conditional waiting times.

The difference between the adaptive memory and Markovian models becomes more evident when the corresponding transition matrices are compared. There is no difference at the first order, thus we will focus on the comparison between $\tau$-order Markovian and adaptive $\tau-$memory models, in the following.

\begin{figure}[h!]
        \includegraphics[width=0.6\textwidth]{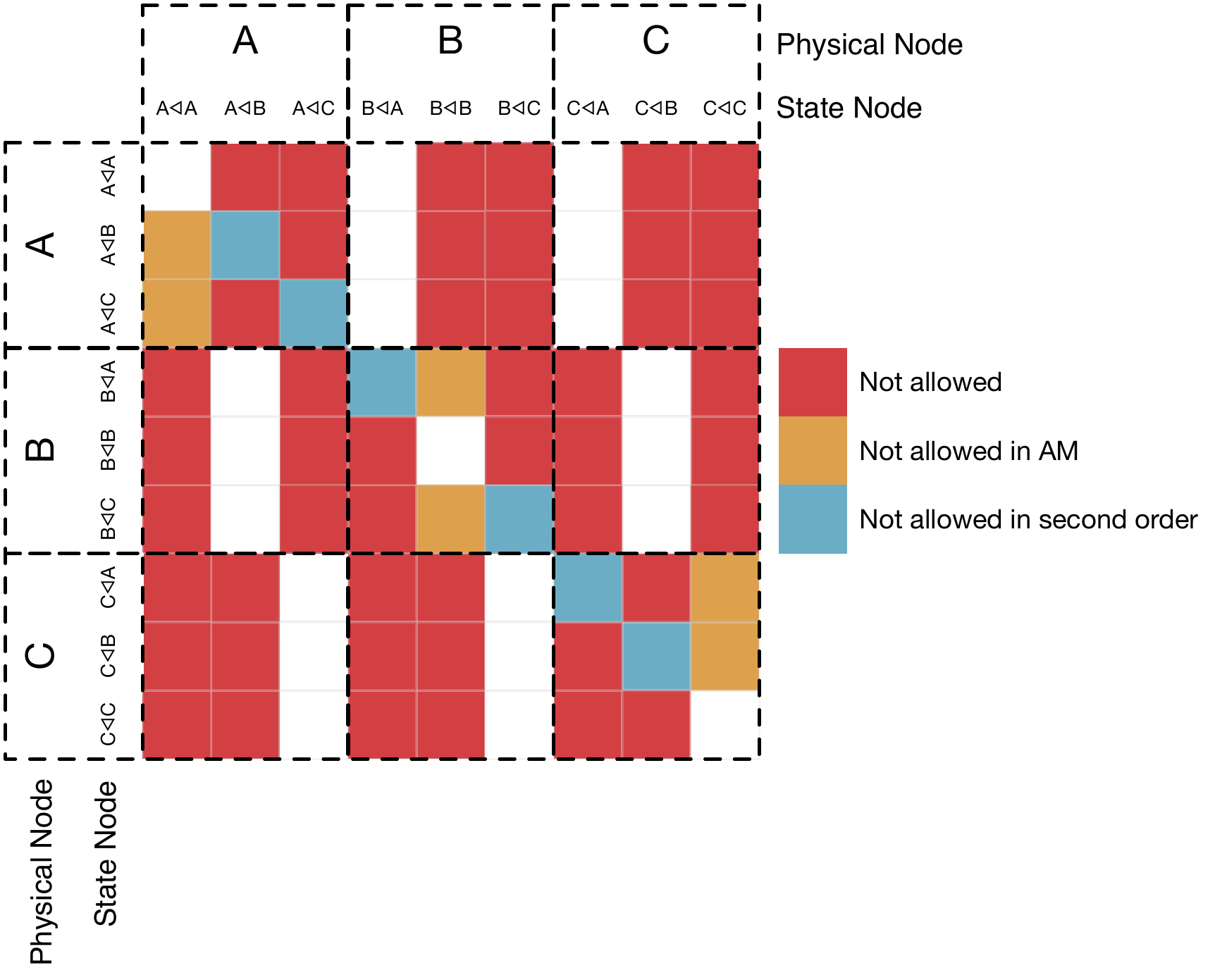}
        \caption{\label{fig:mobilityMatrix}\csentence{Mobility Matrix} Second-order transition matrix for three physical nodes A, B and C. The state nodes are represented with the notation $\bistate{x}{y}$ meaning that walkers in this node have traveled from node $y$ to node $x$. The cells in red are not used by either second-order markovian model or adaptive memory model. The cells in bluish are used only in adaptive memory model, while, the cells in orange are used only in the second-order markovian model. The cells in white are used by both models.}
    \end{figure}

In both models, the number of possible transitions between state nodes is the same and equals $n^{2\cdot(\tau-1)}$, where $n$ is the number of physical nodes. For instance, in second-order models, there are $n^2\times n^2$ transition matrices with $n^3$ possible transitions between state nodes, as shown in Fig.~\ref{fig:mobilityMatrix}. However, the way how each model stores repeating calls in the same physical node is very different. While adaptive memory stores this information into the $n^\tau$ diagonal elements of the matrix, encoding the conditional waiting times discussed in the previous section, Markovian models redistribute this information among off-diagonal entries, because they do not allow this type of self-loops by construction.

More specifically, the information is redistributed among transitions between state nodes of the same physical node. The entries of off-diagonal blocks -- corresponding to transitions between state nodes of different physical nodes -- are the same in both models. Therefore, while the stationary probability of finding a random walker in a physical node is not different in the two models, it is different at the level of state nodes and, as we will see later, this significantly affects diffusion processes such as epidemics spreading.

\subsection{Overview of the dataset}

\begin{figure*}[ht!]
  \centering
  \includegraphics[width=0.8\textwidth]{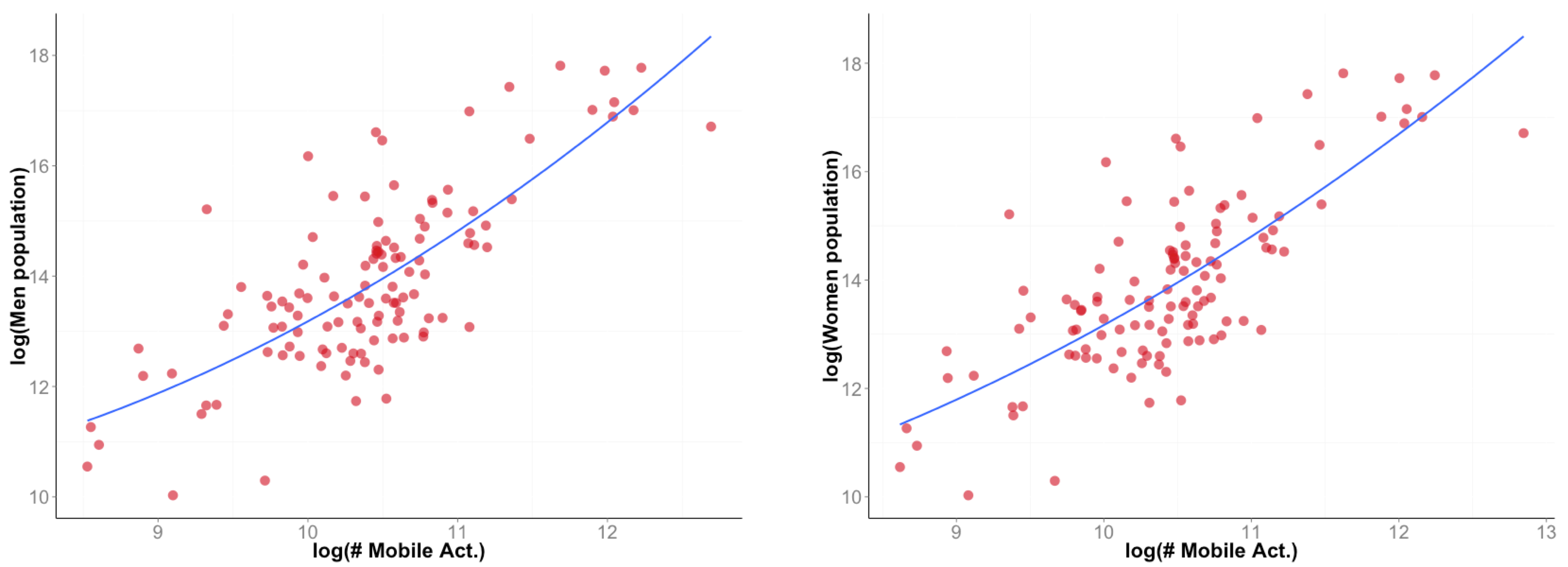}
  \caption{\label{fig:fig4}\textbf{Inferring men and women populations.} Second-order polynomial model (solid line) fitting the log-log relationships between the observed mobile phone data and demographics data (points). Men (A) and women (B) population were fitted separately, thanks to data availability, and have been used to infer the populations in the arrondissements of Bambilor, Thies Sud, Thies Nord, Ndiob and Ngothie.}
\end{figure*}

In the next section, we will quantify the impact of adaptive memory on human mobility modeling by using data sets provided by the Data for Development Challenge 2014 \cite{d4d2014} and some supplementary data sets provided by partners of the challenge. Mobile phone data consist of communications among 1666 towers distributed across Senegal. We exploit this information to map communication patterns between different areas of the country (i.e. the arrondissements). Another subset consists of 560 millions call records of about 150,000 users along one year at at the spatial resolution of arrondissements. We use this information to map individuals' movements among different arrondissements. Demographics information has been obtained from the Senegal data portal~\footnote{\url{http://donnees.ansd.sn/en/}}, an official resource. It is worth noting that information has been manually checked against inconsistencies and data about population for the arrondissements of Bambilor, Thies Sud, Thies Nord, Ndiob and Ngothie were not available. We reconstructed the missing information by combining mobile phone activity and available demographics data, Fig.~\ref{fig:fig4}. We have used the data to infer more realistic contact rates to be used in viral spreading simulations. The contacts among individuals are generally quite difficult to track at country level. Their rate varies depending on several social and demographical factors such as age, gender, location, urban development, \emph{etc.} \cite{rohani2010contact,kiti2014quantifying}. Nevertheless, there are evidences from European and African countries that, on average, the number of daily physical contacts among individuals range from 11 to 22 \cite{rohani2010contact,kiti2014quantifying}. In this work we consider the same contact rate for all arrondissements belonging to the same region. Our choice is justified by the lack of detailed data about the population density per arrondissement. Moreover, we assign 25 contacts per day to the region with highest density (i.e. Dakar) and we rescale the contact rate of all other regions proportionally to the ratio between their density and the density of Dakar, assuming that the minimum contact rate can not be smaller than 10 contacts per day. We obtain a contact rate between 10 and 11 for all regions, except Dakar.

\section{Results}
 
\subsection{Understanding human mobility flow}

\begin{figure}[ht!]
      \includegraphics[width=\textwidth]{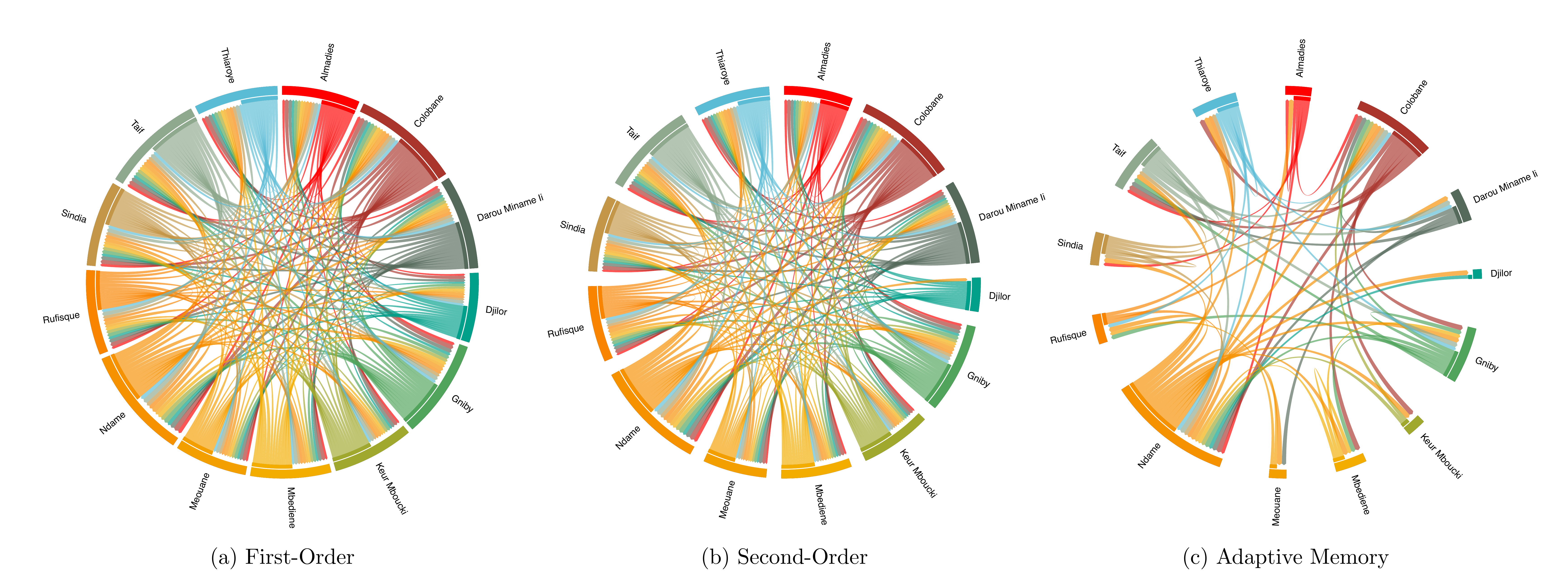}
	    \caption{\label{fig:flow-chart}\csentence{Mobility flow among a sub-set of Senegal's arrondissements}. The figure represents, in logarithmic scale, the modeled mobility flow among a sub-set of arrondissements with the constrain that individuals must pass through Kael (arrondissement in the department of Mbacke, region of Diourbel)  after departing from their origin and before reaching their destination. The figure shows the mobility modeled by means of first-order (A), second-order (B) and adaptive 2-memory (C), putting in evidence the different mobility patterns between Markovian models and adaptive memory. For instance, the adaptive memory module captures returning patterns (i.e. movements like X~$\rightarrow$~Kael~$\rightarrow$~X) better than the first-order model.}
    \end{figure}

We show in Fig.\,\ref{fig:flow-chart} the significant differences in modeling the mobility flow using first-order (FO), second-order (SO) and adaptive memory (AM) models. Markovian models provide very similar transition patterns, whereas adaptive memory provides very different results. The adaptive memory model exhibits significantly less returning transitions than Markovian models, but -- on average -- with much higher probability of observing them. In fact, $47.4$\% of patterns captured by the first-order approach and $43.4$\% captured using second-order are spurious because they are not observed in reality. Remarkably, the probability that an individual comes back to her origin is on average six times higher using adaptive memory models than using first-order, and five times higher using second-order.

To compare the accuracy of both models against the mobility behavior observed in data, we use the coverage, defined as the fraction of nodes visited by an individual within a given amount of time. We calculate the coverage for each individual in the data, over a period of one months, and then we average over all arrondissements to obtain a measure at country level. For the same period of time, we generate three transition matrices $\mathbf{F}$, $\mathbf{H}$ and $\mathbf{A}$ encoding the mobility dynamics for first-order, second-order and adaptive memory models, respectively. To better replicate the calling behavior of the individuals in the data set, we extract information about the distribution of time between calls and we use this information in our simulations (see Appendix~\ref{app:time}). 

    \begin{figure*}[ht!]
        \includegraphics[width=1\textwidth]{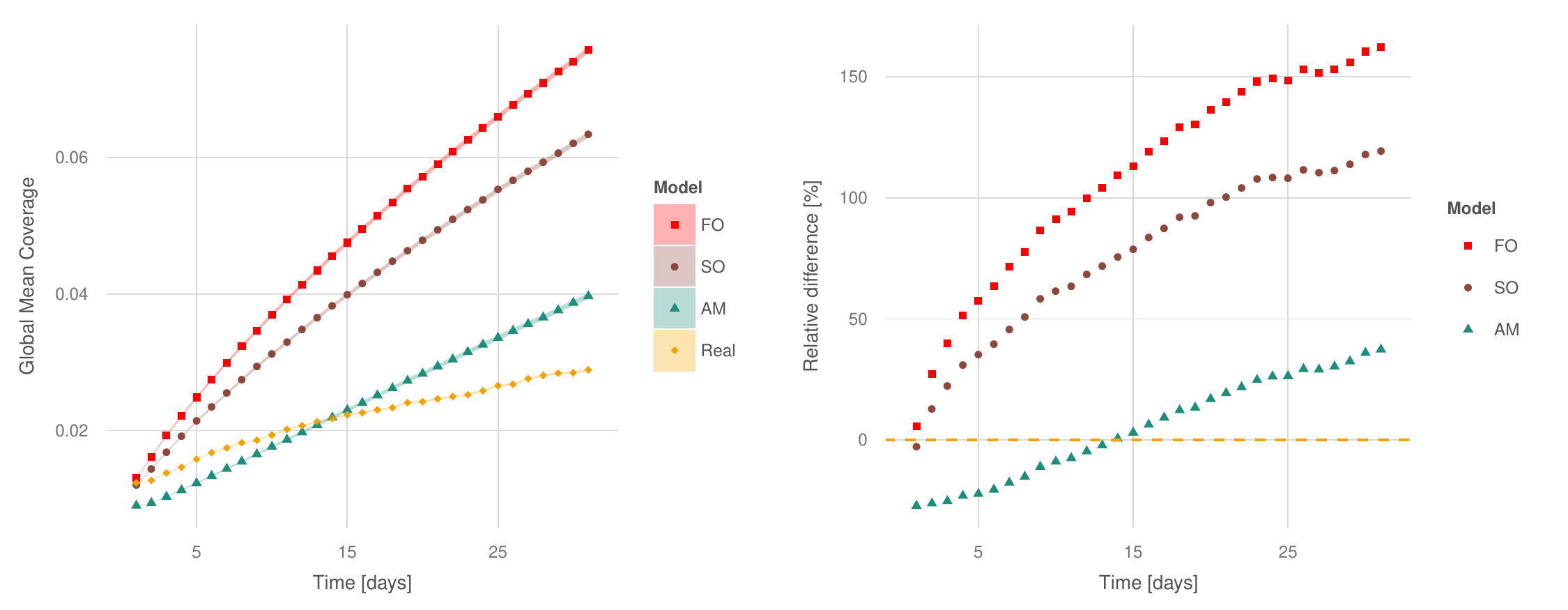}
        \caption{\label{fig:mobilityRes}\csentence{Observed human mobility and theoretical predictions}. (A) Temporal evolution of the global mean coverage calculated from real data and from simulations using first-order (FO), second-order (SO) and adaptive memory (AM) models. (B) Relative difference between the coverage observed in real human mobility and the one obtained from simulations.}
    \end{figure*}

In Fig.~\ref{fig:mobilityRes}A and B we show that people diffuse in the country too fast using Markovian models, whereas significantly slower diffusion is found with adaptive memory, in agreement with empirical observation. These results have deep implications, for instance, in short-term or long-term predictions of epidemic spreading or national infrastructure planning.

\subsection{Impact of human mobility models on the spreading of epidemics}

Here, we focus on epidemic spreading. How infectious individuals move among different locations has a strong influence in how diseases diffuse in a population. We considered each arrondissement as a meta-population where any individual can interact with a limited number of other individuals. We use a SEIR compartmental model~\cite{keeling2008modeling} to characterize the epidemics evolution within each arrondissement and mobility models to simulate people traveling in the country.

The discrete time step of the following models is $\Delta t \approx 1$~hour, approximately the observed median between two successive calls from the same individual. The parameters are demographical and epidemiological. Demographics parameters include the birth $B = \tilde{B}\Delta t$ and death $\delta=\tilde{\delta}\Delta t$ probability, whereas epidemiological parameters correspond to the latent period $\tau_{E}$ of the infection, from which the probability $\epsilon=\Delta t/\tau_{E}$ to pass to the infectious state is calculated, and the infectious period $\tau_{I}$, from which the probability $\gamma=\Delta t/\tau_{I}$ to recover from or die because of the infection is calculated. The last parameter is the effective transmission probability 
\begin{eqnarray}
\beta_{i}(t)=1-\left(1-\tilde{\beta}\Delta t \frac{I_{i}(t)}{N_{i}(t)}\right)^{c_{i}\Delta t},
\end{eqnarray}
an arrondissement-dependent parameter that depends on the average number of contacts per unit of time $c_{i}$ experienced by an individual in node $i$, the fraction of infected individuals in that node and the transmission risk $\tilde{\beta}\Delta t$ in case of contact with an infectious individual. In fact, the definition of $\beta_{i}(t)$ induces a type-II reaction-diffusion dynamics \cite{colizza2007reaction} accounting for the fact that each individual does not interact with \emph{all} the other individuals in the meta-population, but only with a limited sample. If the number of infected agents is small (i.e. $I_{i}(t)\approx0$) the Taylor expansion of $\beta_{i}(t)$ truncated at the first order gives the classical factor $\tilde{\beta}\Delta t c_{i}\Delta t\frac{I_{i}(t)}{N_{i}(t)}$ \cite{keeling2008modeling}. It follows that the equations describing the average spreading of a disease according to a SEIR model coupled to first-order mobility are given by
\begin{eqnarray}
S_{i}(t+1)&=&\sum_{j=1}^{n}F_{ji}\[(1-\delta - \beta_{j}(t))S_{j}(t)+BN_{j}(t)\]\nonumber\\
E_{i}(t+1)&=&\sum_{j=1}^{n}F_{ji}\[ (1-\epsilon-\delta)E_{j}(t)+ \beta_{j}(t) S_{j}(t)\]\nonumber\\
I_{i}(t+1)&=&\sum_{j=1}^{n}F_{ji}\[ (1-\gamma-\delta)I_{j}(t)+\epsilon E_{j}(t) \]\nonumber\\
R_{i}(t+1)&=&\sum_{j=1}^{n}F_{ji}\[ (1-\delta)R_{j}(t)+\gamma I_{j}(t) \]
\end{eqnarray}

    \begin{figure*}[ht!]
        \includegraphics[width=0.9\textwidth]{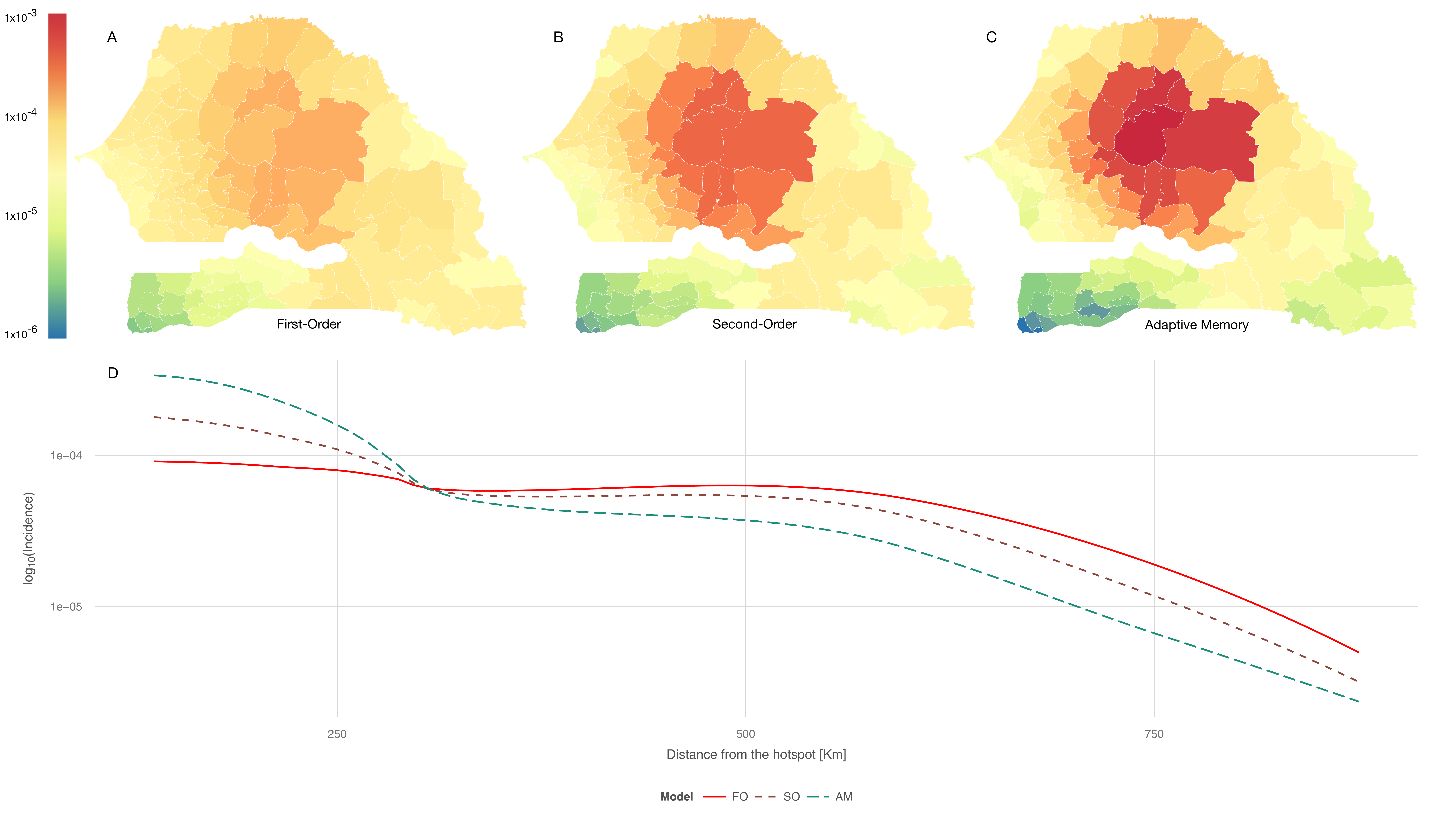}
        \caption{\label{fig:SpreadingFig}\csentence{Spreading of an influenza-like outbreak in Senegal}. We show the incidence of an influenza-like virus over Senegal arrondissements a week after the infection onset, using first-order (A), second-order (B) and adaptive 2-memory (C) mobility models. The infection started in Barkedji (center of Senegal), where three individuals are initially infected. A SEIR compartmental dynamics with parameters $\beta=0.05$,~$\epsilon=0.2$,~$\gamma=0.5$ is used to simulate the spreading of the disease within each arrondissement. We found that the number of arrondissements with infected individuals is higher using Markovian dynamics. Conversely, the adaptive memory favors a higher concentration of infected individuals in the arrondissements around the initial location of the infection. In fact, the location of the onset of the epidemic can be better identified using adaptive memory rather than Markovian models. (D) Relation between the incidence in a region and the distance from the hotspot of the infection using the three models. Adaptive memory models spread the incidence on regions closer to the hotspot and this effect is even more evident when higher memory is used.}
    \end{figure*}

whereas the coupling to the second-order model is given by
\begin{eqnarray}
\tilde{S}_{\alpha}(t+1)&=&\sum_{\psi=1}^{n^{2}}H_{\psi\alpha}\[ (1-\delta-\tilde{\beta}^{(\alpha)}_{\psi}(t))\tilde{S}_{\psi}(t) + B\tilde{N}_{\psi}(t) \]\nonumber\\
\tilde{E}_{\alpha}(t+1) &=& \sum_{\psi=1}^{n^{2}}H_{\psi\alpha}\[ (1-\epsilon-\delta)\tilde{E}_{\psi}(t) + \tilde{\beta}^{(\alpha)}_{\psi}(t) \tilde{S}_{\psi}(t) \]\nonumber\\
\tilde{I}_{\alpha}(t+1) &=& \sum_{\psi=1}^{n^{2}}H_{\psi\alpha}\[ (1-\gamma-\delta) \tilde{I}_{\psi}(t) + \epsilon \tilde{E}_{\psi}(t)\]\nonumber\\
\tilde{R}_{\alpha}(t+1) &=& \sum_{\psi=1}^{n^{2}}H_{\psi\alpha}\[ (1-\delta)\tilde{R}_{\psi}(t) + \gamma\tilde{I}_{\psi}(t) \]\nonumber\\
\tilde{\beta}^{(\alpha)}_{\psi}(t) &=& 1-\left(1-\tilde{\beta}\Delta t\frac{\sum\limits_{\rho=\floor{\frac{\alpha}{n}}n+1}^{\floor{\frac{\alpha}{n}}n+n}\tilde{I}_{\rho}(t)}{\sum\limits_{\rho=\floor{\frac{\alpha}{n}}n+1}^{\floor{\frac{\alpha}{n}}n+n}\tilde{N}_{\rho}(t)}\right)^{c_{i}\Delta t}
\end{eqnarray}
where $N(t)=\sum\limits_{\psi=1}^{n^{\tau}}\tilde{N}_{\psi}(t)$ is the total population in the country at time $t$, $\floor{\cdot}$ indicates the floor function and is used to identify the sub-set of state-nodes corresponding to the same physical node the population $\tilde{S}_{\alpha}$ belongs to. The equations for the adaptive memory model are the same, except that the transition matrix $\mathbf{A}$ is used instead of $\mathbf{H}$.

We initiate the simulation by infecting five individuals in Barkedji, at the center of Senegal. The differences between the diffusion of the infective process using each mobility model are quite visible in Fig.~\ref{fig:SpreadingFig}. The spreading is faster for Markovian models, with some arrondissement populated by more infected individuals than adaptive memory. The incidence, i.e. the fraction of infected individuals in an arrondissement, follows different spatial patterns in the three models (see Fig.~\ref{fig:SpreadingFig}A--C), with a higher incidence observed in the origin of the infection that decreases as we move far from there. This effect is significantly stronger using adaptive memory because it tends to concentrate more infectious individuals close to the origin (see Fig.~\ref{fig:SpreadingFig}D). 
    
\section*{Discussion}

Modeling how people move among different locations is crucial for several applications. Given the scarcity of information about individuals' movements, often human mobility proxies such as call detail records, GPS, \emph{etc}, are used instead. Here, we have shown that dynamical models built from human mobility proxies can be significantly wrong, underestimating (or overestimating) real mobility patterns or predicting spurious movements that are not observed in reality. We have proposed a general solution to this issue, by introducing an adaptive memory modeling of human mobility that better captures observed human dynamics and dramatically reduces spurious patterns with respect to memoryless or higher-order Markovian models. We have validated our model on a data set consisting of 560 millions of call detail records from Senegal. We have found that individuals tend to diffuse faster with standard mobility models than what observed in reality, whereas the adaptive memory approach reconciles empirical observations and theoretical expectations. Our findings have, for instance, a deep impact on predicting how diseases spread in a country. While standard approaches tending to overestimate the geographical incidence of the infection, the more realistic modeling obtained by means of adaptive memory can improve the inference of the hotspot of the infection, helping to design better countermeasures, e.g. more effectives quarantine zones, improved resources deployment or targeted information campaigns.


\section*{Competing interests}
The authors declare that they have no competing interests.

\section*{Data Accessibility}
The data used in this manuscript were made publicly available during the D4D Senegal Challenge organized by Orange. More information about data can be found here: \url{http://www.d4d.orange.com/en/presentation/data}

\section*{Author's contributions}
JTM and MDD developed the theoretical model and carried out the statistical analyses; MDD and AA conceived of the study, designed the study and coordinated the study. All authors wrote the manuscript and gave final approval for publication.

\section*{Funding}

M.D.D. acknowledges financial support from the Spanish program Juan de la Cierva (IJCI-2014-20225). JTM was supported by Generalitat de Catalunya (FI-DGR 2015). A.A. acknowledges financial support from ICREA Academia and James S.\ McDonnell Foundation and Spanish MINECO FIS2015-71582. 

\bibliographystyle{apsrev} 
\bibliography{adaptive}      

\newpage
\appendix

\section{Accounting for bursty behavior in the calculation of the coverage}\label{app:time}

\begin{figure*}[!t]
  \centering
  \includegraphics[width=1\textwidth]{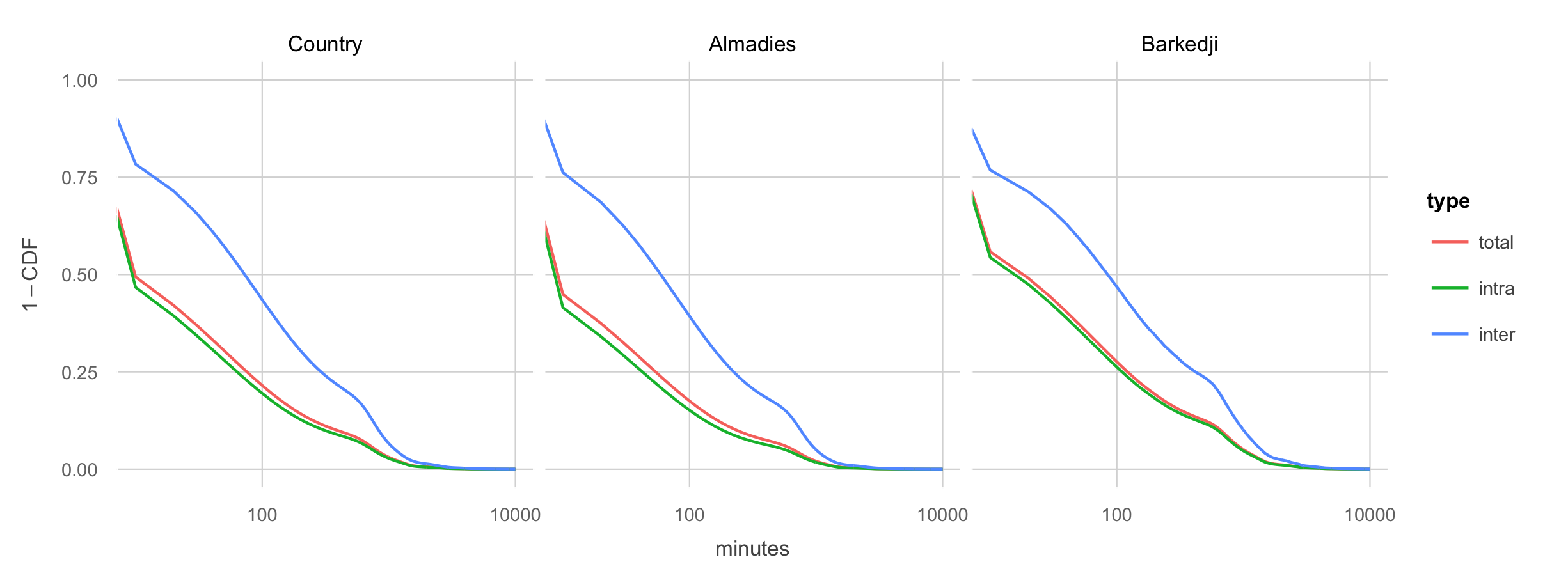}
  \caption{Complementary cumulative density function of time between calls at country level and in Almadies and Barkedji arrondissements. For each of them we consider three different actions: time between sequential calls in the same arrondissement ($P_{intra}$), time between consecutive calls performed in different arrondissements ($P_{inter}$), and time between any calls ($P_{total}$). We observe how using information at country level is not a good proxy for modeling the dynamics because the behaviour is significantly different depending on the arrondissement. The time between calls performed in Almadies is lower than in Barkedji due the high density of arrondissements in the region of Almadies, reducing the traveling time between different locations. Note that scale is logarithmic on the $x-$axis.} 
  \label{fig:fig6}
\end{figure*}

We found that human communications are bursty, with bunches of intense activities concentrated in a short period that are separated by longer periods of inactivity, Fig.~\ref{fig:fig6}. This pattern is evident when two consecutive calls are made in the same arrondissement. However, in the case of two consecutive calls made from different arrondissements, the probability is lower because of the time required to move from one place to another. The arrondissement where the individual lives influences these activities: e.g. the Dakar region consists of several small arrondissements that are close together and the time required for movements is smaller when compared with the time observed for people living in Louga, a more sparse region. We considered these empirical facts in our simulations, by extracting two call distributions for each arrondissement, $P(T_{intra})$ and $P(T_{inter})$, to model the dynamics within the same arrondissement and among different arrondissements, respectively. An independent simulation is performed using each arrondissement as the starting point of a traveller's journey. At each time step, the traveller moves according to the transition probabilities of a specific mobility model. If the traveller decides to stay in the same arrondissement, the elapsed time within the simulation is increased by a random number drawn from the distribution $P(T_{in})$. Otherwise, if the traveller decides to change her location, the elapsed time is increased by a random number drawn from distribution $P(T_{out})$. The traveller stops her journey when the elapsed time reaches 30 days. The coverage for each arrondissement is computed as the average of five thousand independent realizations of this process.

\end{document}